\documentclass[12pt]{article}

\usepackage{scicite}
\newcommand\ddfrac[2]{\frac{\displaystyle #1}{\displaystyle #2}}

\usepackage{dirtytalk}
\usepackage{graphicx}
\graphicspath{{images/} }
\usepackage{float}
\usepackage{times}
\usepackage{lineno}
\usepackage[normalem]{ulem}
\usepackage{xcolor}
\usepackage{hyperref}

\title{The Paradox of Second-Order Homophily in Networks}

\author
{Anna Evtushenko,$^{1, 3\ast}$ Jon Kleinberg$^{1, 2, 3}$\\
\\
\normalsize{$^{1}$Department of Information Science}\\
\normalsize{$^{2}$Department of Computer Science}\\
\normalsize{$^{3}$Cornell University, Ithaca NY, USA}\\
\\
\normalsize{$^\ast$anna@infosci.cornell.edu}
}

\date{}

\begin{document}

\baselineskip24pt

\maketitle

\begin{abstract}
Homophily---the tendency of nodes to connect to others of the same
type---is a central issue in the study of networks.
Here we take a local view of homophily, defining notions of first-order
homophily of a node (its individual tendency to link to similar others)
and second-order homophily of a node (the aggregate first-order homophily
of its neighbors).
Through this view, we find a surprising result for homophily values
that applies with only minimal assumptions
on the graph topology. It can be phrased most simply as \say{in a graph of red and blue nodes, red friends of red nodes are on average more homophilous than red friends of blue nodes.}
This gap in averages defies simple intuitive explanations, applies to globally heterophilous and homophilous networks and is reminiscent of but structurally distinct from the Friendship Paradox. The existence of this gap suggests intrinsic biases in homophily measurements between groups, and hence is relevant to empirical studies of homophily in networks.
\end{abstract}

\section{Introduction}
Homophily---the principle that nodes in a network tend to
have links to other nodes with similar attributes---is one of
the most robust principles governing real-world network structure.
The term originates in the study of social connections among people,
where an increased tendency to link between people of the same gender \cite{stehle2013gender},
race \cite{moody2001race}, ethnicity \cite{qian2007social}, religion \cite{cheadle2012friendship}, 
or social status \cite{mcpherson1987homophily} has
been extensively studied \cite{currarini2009economic,kossinets2009origins,mcpherson-homophily,smith2014social,karimi2018homophily,lee2019homophily,avin2020mixed,asikainen2020cumulative}.
More broadly, it has been shown to be a powerful framework for networks
in a wide range of domains, with the level of {\it assortativity}
\cite{newman-assortativity,newman-mixing-patterns}---the correlation of attributes across links---serving as
a quantity of interest for many applications.

Homophily has traditionally been discussed as a global property of a network.
But a powerful idea in analyzing complex real-world phenomena is disaggregating global properties to explore how a
macro-level signal arises from a composite of individual-level attributes
\cite{granovetter-threshold,schelling-micromotives}.
This has been a particular focus in studies of how aggregate network properties
can arise from local phenomena
\cite{barabasi2016network,easley-kleinberg-ncm,jackson-networks-book,newman-networks-book}.
If we apply this principle to homophily, we can study the 
{\it local homophily} at each individual node in a network,
capturing that node's tendency to link
to similar neighbors \cite{altenburger2018monophily}:
formally, in a network where nodes may belong to one of multiple
types, we say that the local homophily at $i$ is the fraction
of $i$'s neighbors that are of the same type as $i$.
Much as the {\it degree} of a node $i$ in a network
represents a local property of the network's overall connectivity,
specialized to $i$,
our local notion of homophily at a node $i$ will analogously represent the 
extent to which the overall similarity across links is manifested
in the vicinity of $i$.

With any local parameter measured at each node of a network,
we can study how this parameter varies across the network.
Typically this investigation is carried out empirically, since
patterns of variation are generally domain-specific.
But there are occasional, surprising instances in which the 
pattern of variation is governed by a purely mathematical constraint---one which applies across all networks, with only minimal assumptions on graph topology.
This is true for the case of node degrees, which are governed by
a mathematical result known as the {\it Friendship Paradox}---the principle that \say{your friends have more friends than you do}
\cite{feld-friendship-paradox,kramer2016multistep}.
Since its discovery roughly forty years ago, the Friendship Paradox
for node degrees has spurred similar findings for other node attributes such as eigenvector centrality or close correlates of degree
\cite{lerman2016majority,eom2014generalized,jackson2019friendship,ugander2011anatomy,higham2019centrality}---results in a \say{me versus my friends} genre that compare the value of a node's attribute to the value of this attribute at the node's neighbors.
But there have been no rigorous results in what we might call the 
\say{friends of one type versus friends of another type} genre, in which we start with a network containing two types of nodes (as is necessary to define a meaningful notion of homophily for example) and compare an attribute at the friends of one type of node to the value of this attribute at the friends of a different type of node.

In this work, we use our local view of homophily to identify 
a novel, \say{paradoxical} constraint on the pattern of local homophily values,
which holds across essentially all undirected network topologies{--- homophilous \textit{and} heterophilous in the global sense of the word---}in which nodes can come from one of two types. (Like the Friendship Paradox, it requires only a minimal non-degeneracy
assumption on the network.)
We will define the \say{Homophily Paradox} precisely in what follows, but to convey the idea
informally, we describe it here in terms of a simple social-network example.
Imagine an undirected social network at a school with students who come from two
different years; we will call them {\it junior students} and 
{\it senior students}.
We will think of these as distinct types of nodes in the social network;
in this way, 
the local homophily of a node is the fraction of neighbors who are
students from the same year.
Since we are interested in the effects of variation in the homophily values, we will make the following minimal assumption: 
that the local homophily values
among the students of each type are not all the same. Note that this means there is at least one senior-senior edge---since otherwise the senior nodes would all have local homophily 0---and at least one junior-senior edge---since otherwise the senior nodes would all have local homophily 1. In particular, this implies, for example, that the underlying network is not a bipartite graph with the senior and junior nodes forming the two sides of the bipartition.
Now suppose we list all the senior friends of each senior student, and
merge these lists (counting people multiple times 
when they appear in multiple lists)
to create a single list of the senior friends of senior nodes.
We now do the same thing for the senior friends of each junior student,
creating a second merged list, this one of the senior friends of junior nodes.
(Notice that because there is at least one senior-senior edge and one junior-senior edge---a consequence of homophily values of the senior nodes not being all the same---these two lists are both non-empty.)
Our finding is the following: in a network like the above, the average of the local homophily values is strictly larger in the first list than it is 
in the second. In other words, senior friends of senior nodes are more homophilous than senior friends of junior nodes.

The Friendship Paradox has an interpretation in terms of the perceptions of nodes---that people tend to assume the degrees of others are not higher than their own, but they are---a counter-intuitive finding. Our result can also be interpreted in terms of local perceptions in a network, but in our case it concerns how two groups form different perceptions of homophily values in the same network, and the mathematical result arguably supports a possible intuition. For example, if the two types in the network were Democrats and Republicans, then Democrats would plausibly think of their Democratic friends as more homophilous than Republicans would think of \emph{their} Democratic friends---intuitively, perhaps, because a Republican individual would have a clear example (in themselves) that their Democratic friends are not strictly party-aligned when it comes to friendship. This distinction in homophily values---comparing the Democratic friends of Democrats to the Democratic friends of Republicans---is what our mathematical result demonstrates in a general setting.

Our result does not require any assumptions on the structure of
the network other than the minimal condition described above---that there is diversity in the local homophily values, which in turns implies a second requirement that the two lists in question are not empty so their means can be defined.
As such, the result is not an empirical statement about how homophily values are structured 
in any particular domain, but about the intrinsic nature of homophily itself.
Indeed, it does not even require that nodes preferentially
link to their own type; it continues to hold even if nodes tend to link mainly to the other type (in a globally heterophilous network).

The intuition for why the result is true can be found in the definition of homophily
itself:
the homophily diversity in the senior nodes (in our example)
tells us that senior nodes who have higher local homophily values, and therefore contribute more to the average, will be counted more times by their more numerous senior friends and fewer times by their less-numerous junior friends. 
As with the Friendship Paradox, which similarly has a brief conceptual justification, the result is only \say{paradoxical} in the sense that
it requires some careful reconciliation with its initial statement.

In the remainder of the paper, we provide 
the formal definitions underlying the result, 
followed by its precise statement and proof.
We also note some further analogies to the Friendship Paradox.
In particular, the mild assumptions underlying our result are
similar to those in the traditional statement of the Friendship Paradox,
which requires that
the degrees not all be the same in order to have a strict inequality
and which tends to exclude nodes of zero degree from consideration.
Moreover, like the Friendship Paradox, our result has a formulation
both in terms of the enumeration of a list of nodes 
(the general result described above)
as well as in terms of a global average of average values at each node
(which we formalize in what follows, check empirically,
and prove in a special case).
However, despite these analogies, there is a strong sense in which 
our result does not follow from the Friendship Paradox, nor is it
even the direct analog of it for the notion of homophily;
as we also show in what follows, a direct attempt to mirror the
Friendship Paradox for the case of homophily, such as a principle that \say{your friends are more homophilous
than you are,} leads to statements that are false in general.
The formulation in terms of lists of neighbors of different types
seems crucial to obtaining an inequality that applies in general networks.

\section{Definitions and Statement of Main Result}

\begin{figure}[H]
\centering
\includegraphics[width=0.5\textwidth]{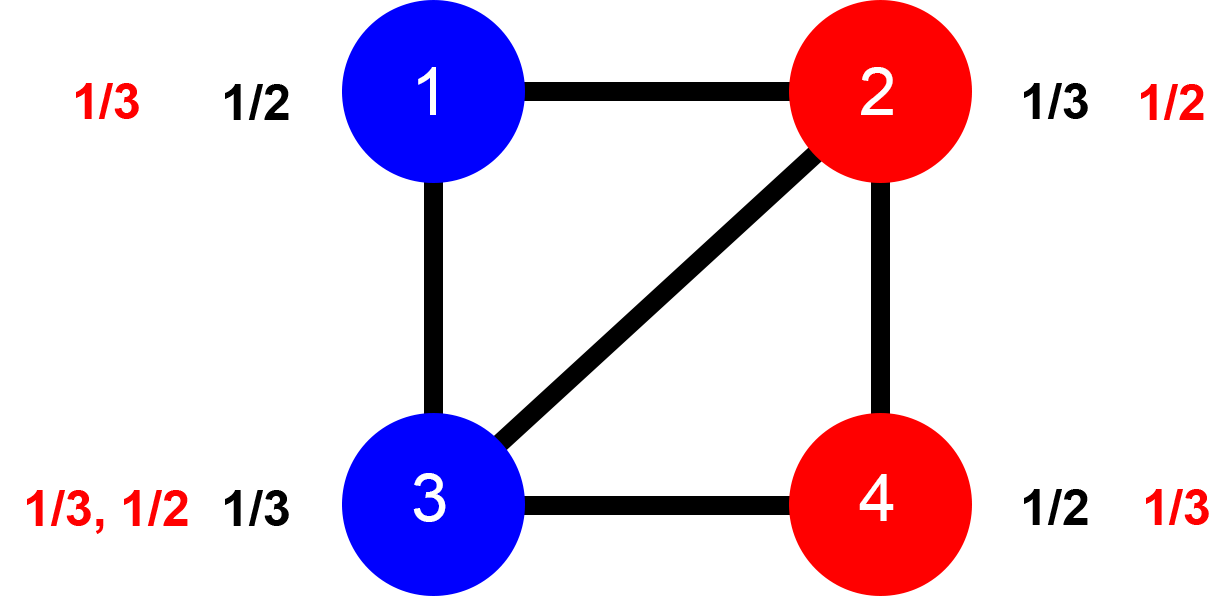}
\caption{Defining first-order and second-order homophily in a network. Here, the first-order homophily of each node is their proportion of friends of the same type, shown in black. The second-order \textit{red} homophily of each node is either the list or the mean of the first-order homophily values of the node's red neighbors. The list version of second-order red homophily of each node is shown in red.
\label{fig:example}
}
\end{figure}

We consider networks in which each node belongs
to one of two types, and we define the \textit{first-order homophily},
or \textit{local homophily}, of a node $i$ (also sometimes referred to as a \textit{seed}) to be the fraction of 
$i$'s neighbors that are of the same type as $i${, denoted $h_i$}.
There are many contexts in which nodes might be of one of two types---for example, any setting with a property such that some nodes exhibit
the property (forming one type) and other nodes don't exhibit the property
(forming the other type).
For simplicity, we will refer to the types of nodes as \textit{red} and 
\textit{blue} in what follows.
Figure \ref{fig:example} depicts an example, with the first-order
homophily of each node written in black.

We will think of an aggregation of
the first-order homophily values of
$i$'s neighbors as the \textit{second-order homophily} of $i$.
In the main definition we consider, we take node $i$'s 
second-order homophily to be the \textit{list} of first-order homophily values of
its neighbors.
We will be particularly interested in the second-order homophily not
just over all of $i$'s neighbors, but over just the red neighbors
(the second-order red homophily of $i${, $s^{(R)}_i$}) 
and just the blue neighbors of $i$
(the second-order blue homophily of $i${, $s^{(B)}_i$}){, since in only looking at a certain type of friends of $i$ we can more meaningfully compare their first-order homophily values}. 
Note that for the main definition it's okay for some of these lists to be empty if a node has no friends of a certain type. At the same time, as mentioned above, at least one node of each type should have a non-empty list for the global averages, which are the basis of the result, to be defined. This statement is equivalent to there being a red-red and a blue-red edge if we're looking at red second-order homophily values $s^{(R)}_i$. Figure \ref{fig:example} shows $s^{(R)}_i$ of
each node $i$ written as a list in red. 

Finally, we list some additional definitions that will be useful in what follows. The \textit{mean first-order homophily of red nodes}{, $\lambda_{R}$,} is the mean of first-order homophily values of red nodes. The \textit{mean second-order red homophily of red nodes}{, $\mu^{(R)}_{R}$,} is derived from concatenating all the second-order red homophily values of red nodes (in list form) and taking the mean of this combined list. Any instances of \say{red} in these definitions may be replaced with \say{blue,} giving us 2 first-order means and 4 second-order means. \textit{Homophily diversity} for nodes of a type is the standard deviation of the first-order homophily of nodes of that type (denoted $\sigma_{R}$ for red nodes, $\sigma_{B}$ for blue nodes). We say that \textit{there is} homophily diversity in red nodes if the standard deviation of first-order homophily of red nodes is not 0{, or if $\sigma_{R}>0$}. The same can be said for blue nodes.

\paragraph{A Gap Phenomenon.}
Our main result is a fundamental gap phenomenon that applies
to second-order red homophily in any undirected network with homophily diversity in red nodes:
the mean second-order red homophily of red nodes ({$\mu^{(R)}_{R}$}) is larger than
the mean second-order red homophily of blue nodes ({$\mu^{(R)}_{B}$}). We define this difference, $\mu^{(R)}_{R}-\mu^{(R)}_{B}$, as a red gap $g^{(R)}$, and prove that it is positive below. Note that if the graph does not exhibit homophily diversity in red nodes, such that all red nodes have homophily $h$, but there is still at least one red-red and one blue-red edge, we must have 
$\mu^{(R)}_{R}=\mu^{(R)}_{B} = h$, since each mean is an average of some sample of first-order homophily values of red nodes, all of which are $h$. If there isn't a red-red or a blue-red edge, at least one of $\mu^{(R)}_{R},\mu^{(R)}_{B}$ is undefined.
By the symmetry of red and blue in all our definitions, the
analogous set of statements holds for second-order blue homophily ($g^{(B)}=\mu^{(B)}_{B}-\mu^{(B)}_{R}>0$ iff $\sigma_B>0$ and there is a blue-blue and and red-blue edge);
for ease of exposition, we will discuss the results in terms of second-order red homophily only.

As an example of the positive gap, in Figure \ref{fig:example}, 
the concatenation of the second-order red homophily values of the red
nodes yields the list $\frac{1}{3}, \frac{1}{2}$, with a mean of 
${\mu^{(R)}_{R}} = \frac{5}{12}$, while 
the concatenation of the second-order red homophily values of the blue
nodes yields the list $\frac{1}{3}, \frac{1}{3}, \frac{1}{2}$, with a mean of 
${\mu^{(R)}_{B}} = \frac{7}{18} < \mu^{(R)}_{R}$ giving $g^{(R)}>0$.

{Since the result has a colloquial formulation of \say{red friends of red nodes are on average more homophilous than red friends of blue nodes}, it may lead to an \say{intuitive} explanation that this is due to the fact that while all red friends in question have a red friend as per our assumption, red friends of blue nodes also definitely have a blue friend, which leads to their overall lower homophily in the aggregate. But had this been correct, the same reasoning would apply for the case with constant local homophily for the red nodes, whereas as we have shown, zero red homophily diversity leads to zero gap in second-order homophily means.}

The positive-gap result is not dependent on empirical assumptions
about network data, or on probabilistic assumptions about
a generative model for networks; it holds in all graphs satisfying the constraint above.
But it has empirical consequences: to the extent that we seek
to learn about the value of homophily at each node in a network individually,
there are inherent biases in certain kinds of values 
(such as the mean second-order red homophily) as we observe them between 
different types of nodes. 
\paragraph{The Singular Version.}
There are other natural ways to aggregate the first-order homophily values
of a node $i$ so as to produce a meaningful notion of second-order homophily.
An alternative aggregation we consider is taking the mean of the list of first-order homophily values of $i$'s neighbors.
(As before, we can also consider this for just the red neighbors
or just the blue neighbors of $i$.)
We will refer to this mean number as the \textit{singular} version of second-order homophily, $s^{(R\textrm{,sing})}_i$ or $s^{(B\textrm{,sing})}_i$,
because it produces a single value, in contrast to the \textit{list} form
of second-order homophily defined previously (which produces a
list of values for each node $i$). 
So, if the list version of second-order red homophily of $i$ is a list $s^{(R)}_i$, the singular version of second-order red homophily of $i$ is the mean of $s^{(R)}_i$. Since every red (or blue) node's second-order red homophily $s^{(R\textrm{,sing})}_i$ is now a number, the mean second-order red homophily of red (or blue) nodes would be simply the mean of those numbers, denoted $\mu^{(R\textrm{,sing})}_{R}$ or $\mu^{(R\textrm{,sing})}_{B}$ respectively. Note that in order to be able to define $s^{(R)}_i$ for each node, we need to avoid zeros in denominators and further require that each node has at least one red friend (and/or one blue friend, \textit{if} we are interested in the blue gap).
When we study this property in empirical network data, we prune nodes that do not satisfy this property; very few nodes are eliminated by this requirement in practice.

In Figure \ref{fig:example}, the list version of second-order red homophily of node 3, $s^{(R)}_3$, is the list $\frac{1}{3}, \frac{1}{2}$, and the
singular version $s^{(R\textrm{,sing})}_3$ for this same node is $(\frac{1}{3}+\frac{1}{2})/2=\frac{5}{12}$. Other values of second-order red homophily remain the same between the two versions. The red gap $g^{(R\textrm{,sing})}$ is still positive in this singular version world, as $(\frac{1}{3}+\frac{5}{12})/2=\frac{3}{8}={\mu^{(R\textrm{,sing})}_{B}}<{\mu^{(R\textrm{,sing})}_{R}} = \frac{5}{12} = (\frac{1}{2}+\frac{1}{3})/2 $. 
The contrast between the list and singular versions of our gap results is reminiscent of similar contrasts between taking a single global average or an average of averages.
For example, this contrast arises in the definition of the clustering coefficient for networks. 
The global clustering coefficient of a network is calculated as a fraction of all closed triplets over all closed or open triplets---one overall fraction, much like the list version of second-order red homophily where the mean is calculated once. The average clustering coefficient, on the other hand, is the mean of local clustering coefficients, each of which is a fraction---this coefficient is a fraction with fractions in the numerator---much like the singular version of the mean second-order red homophily where each node’s contribution is a fraction, and the mean is the result of dividing the sum of all the fractions by their quantity.

\paragraph{Overview.}
We now describe the plan for the remainder of the paper.

We first prove the main result, a positive gap for
the list version of second-order red homophily.
We also discuss the singular version of second-order red homophily;
we conjecture the gap to be positive in general for this case
as well (as long as all first-order homophily values of red nodes are not identical, if talking about the red gap),
and prove this for a special case.

The fact that the gap is positive does not directly tell us how
large it typically is.
We examine the gap size both in empirical data and in synthetic probabilistic
models, finding that it is non-trivially large in both.
For empirical data, we consider the social networks in the
Facebook100 dataset \cite{traud2012social}, using reported gender as the type.
We also derive analytical formulas for the gap size in 
some basic random-graph models.

Finally, we consider connections and extensions of the work.
We explore the connection to the Friendship Paradox as described above,
showing how a more direct attempt at formulating a closer analogy
to the Friendship Paradox for homophily produces a statement that
is in fact false in general, though true for graphs where all
nodes have the same degree.
We also conclude by discussing generalizations of the definitions
to the case of more than two types, as well as possible applications.

\section{The List Version of Second-Order Homophily}

We now provide a high-level description of the proof of
the positive gap for the list version of second-order red homophily.
See Section 1 in the Supplementary Information for the complete step-by-step proof.

Suppose there are $k$ red nodes, and the $i^{\rm th}$ red node has
homophily $h_i$ and degree $d_i$.
We require $\sigma_R>0$, meaning the homophily values $h_1, h_2, \ldots, h_k$ are not all equal. Each red node $i$ with homophily $h_i$ and degree $d_i$ has
$h_i d_i$ red friends and $(1 - h_i) d_i$ blue friends. Hence for each red node $i$, their homophily $h_i$ will be counted $h_i d_i$ times in $\mu^{(R)}_{R}$, the mean second-order red homophily of red nodes, and $(1 - h_i) d_i$ times in $\mu^{(R)}_{B}$, the mean second-order red homophily of blue nodes. These quantities would thus be given by

\begin{equation}
{\mu^{(R)}_{R}} = \ddfrac{\sum_{i=1}^{k} d_i h_{i}^2 }{\sum_{i=1}^{k} d_i h_{i}} \mbox{and } \mu^{(R)}_{B} = \ddfrac{\sum_{j=1}^{k} d_j (1-h_j) h_{j} }{\sum_{j=1}^{k} d_j (1-h_{j})}
\end{equation}

We'd like to prove that ${g^{(R)} = \mu^{(R)}_{R}-\mu^{(R)}_{B}}>0$.
Bringing the above to the common denominator and multiplying by this positive denominator, rearranging and finding common terms we now need to prove $\sum_{i=1}^{k} \sum_{j=1}^{k} d_i d_j h_{i} (1-h_{j}) (h_i - h_j) > 0$.

First note that if $h_i=h_j$ the sum term is $0$. Then note that if $h_i \neq h_j$, at least one of the terms for $(i,j)$ and $(j,i)$ is not zero. We know from homophily diversity in red nodes that there is at least one pair $(i,j)$ such that $h_i \neq h_j$. Next, we'd like to prove that for all such pairs $(i,j)$, the sum of the two terms $T_{ij} = d_i d_j h_i (1-h_j) (h_i - h_j)$ and $T_{ji} = d_j d_i h_j (1-h_i) (h_j - h_i)$ is positive. $d_i$ and $d_j > 0$ so we only need to prove $(h_i (1-h_j) - h_j (1-h_i)) (h_i - h_j) > 0 $. Suppose $h_i > h_j$ so $h_i - h_j > 0$. $ h_i (1-h_j) - h_j (1-h_i) > 0$ hence $h_i - h_j > 0 $. This is true. Now suppose $h_i < h_j$ so $h_i - h_j < 0$. $ h_i (1-h_j) - h_j (1-h_i) < 0$ hence $h_i - h_j < 0 $. This is true. Since terms for $h_i = h_j$ are $0$ and pair-terms for $h_i \neq h_j$ exist and are positive, the whole sum is positive, so $g^{(R)}>0$ if $\sigma_R>0$.

\section{Proof of the Special Case of the Singular Version}
\label{section:single}

\def\altmu{\mu}

In this section, we consider the alternate way of aggregating
neighbor homophily values in order to define second-order homophily:
the \textit{singular version} in which the second-order homophily of $i$, $s^{R\textrm{,sing}}_i$, is the mean of the first-order homophily values of $i$'s neighbors, as opposed to the list.
We give an overview of our analysis for this case, and
provide the complete details in Section 2 of the Supplementary Information. The special case we study here is admittedly quite constrained, and a later section on gap in empirical data provides further analysis, showing that the gap is positive in real-world data, with Facebook100 data as an example.

A first ingredient in the analysis is the following 
\textit{balance equation}, which 
will be useful at several points. 
Let $R$ be the set of red nodes and $B$ the set of blue nodes;
let the homophily and degree of a red node $i$ be denoted
$h_i$ and $d_i$ respectively, and 
let the homophily and degree of a blue node $j$ be denoted
$p_j$ and $c_j$ respectively.
Then we have the following equality:

\begin{equation}\label{eqn:balance}
\sum_{i \in R} d_i (1-h_i) = \sum_{j \in B} c_j (1-p_j)
\end{equation}

This identity follows by counting the number of undirected edges
with one red and one blue end in two different ways: once from
the red endpoints of the edges (left side of the equation), and once from the blue endpoints of the edges (right side).
As before, each red node $i$ with homophily $h_i$ and degree $d_i$ has
$h_i d_i$ red friends and hence $(1 - h_i) d_i$ blue friends; the sum
over all red nodes gives us the quantity on the left. Similarly,
summing over all blue nodes gives us the quantity on the right.
Since the graph is undirected, 
these are two different ways of counting the same set of 
edges, and so they must be equal.

Now, let $R$ have $k$ red nodes and $B$ have $n-k$ blue
nodes where $n$ is the total number of nodes in a network (later, we also use $r=\frac{k}{n}$ to denote proportion of red nodes). Let $N(i)$ denote the neighbors of a node $i$, and for two sets
of nodes, let $E(S,T)$ be the set of edges with one end in $S$ and the
other end in $T$.

By analogy with the previous section, we would like to consider
the mean second-order red homophily of red nodes and the mean
second-order red homophily of blue nodes, but for the singular version.
We will show that the gap $\mu^{(R\textrm{,sing})}_R - \mu^{(R\textrm{,sing})}_B$ is positive in
the following special case: all red nodes have degree $d$;
all blue nodes have degree $c$ and
all blue nodes have homophily $p$.
The only constraint on first-order homophily of red nodes is $\sigma_R>0$, as before. As stated previously, in order to be able to define second-order red homophily for each node, we further require that each node have at least one red friend.

In this case, 
the mean second-order red homophily for red nodes is:

\begin{equation}
\label{eqn:singular}
{\mu^{(R\textrm{,sing})}_R}= \frac{1}{k} \sum_{i \in R} \left( h_i \sum_{j \in R \cap N(i)} \frac{1}{h_j d}\right)= \frac{1}{k} \sum_{\{i,j\} \in E(R,R)} \left(\frac{h_j}{h_i d} + \frac{h_i}{h_j d}\right) 
\end{equation}

And the mean second-order red homophily for blue nodes is:
\begin{equation}
{\mu^{(R\textrm{,sing})}_B} = \frac{1}{n-k} \sum_{i \in R} \left( h_i \sum_{j \in B \cap N(i)} \frac{1}{(1 - p) c} \right)
\end{equation}

Let's first prove that ${\mu^{(R\textrm{,sing})}_R} > \frac{1}{k} \sum_{i \in R} h_i = \lambda_{R}$. ${\mu^{(R\textrm{,sing})}_R} = \frac{1}{k d} \sum_{\{i,j\} \in E(R,R)} \left(\frac{h_j}{h_i} + \frac{h_i}{h_j}\right) > \frac{1}{k d} \sum_{\{i,j\} \in E(R,R)} 2$. The inequality for reciprocals can be strict because there is homophily diversity in red nodes. Notice that the number of red edges $|E(R,R)|$ is $\frac{1}{2} \sum_{i \in R} h_i d$. Then $\frac{1}{k d} \sum_{\{i,j\} \in E(R,R)} 2 = \frac{1}{k d} \left( \frac{1}{2} \sum_{i \in R} h_i d \right) 2 = \frac{1}{k} \sum_{i \in R} h_i = \lambda_{R}$. Thus ${\mu^{(R\textrm{,sing})}_R} > \lambda_{R}$ in this special case. From the balance equation (see equation~\ref{eqn:balance}), we derive that $p = \frac{cn - (c+d)k + dk \lambda_{R}}{c(n-k)}$. Thus ${\mu^{(R\textrm{,sing})}_B} = \frac{1}{k(1 - \lambda_{R})} \sum_{i \in B} h_i (1-h_i) $. We already know ${\mu^{(R\textrm{,sing})}_R} > \lambda_{R}$. To prove ${\mu^{(R\textrm{,sing})}_R} > \mu^{(R\textrm{,sing})}_B$, we need to show
${\lambda_{R}} \geq \mu^{(R\textrm{,sing})}_B$. Multiplying by $(1 - \lambda_{R})$ and expanding ${\lambda_{R}}$ we get,
$$\left( \frac{1}{k} \sum_{i \in R} h_i \right) \left( \frac{1}{k} \sum_{i \in R} (1 - h_i) \right) \geq \frac{1}{k} \sum_{i \in B} h_i (1-h_i) $$
Now, we can relabel $h$'s so that $h_i \geq h_j$ if $i > j$. This means that $(1 - h_i) \leq (1 - h_j)$ for $i > j$. Now, the last statement is true by (reverse) Chebyshev's sum inequality.

\section{Gap in Empirical Data}
\label{section:data}

The results thus far show that the red gap is always positive
for a graph that exhibits homophily diversity in red nodes, but it is less clear
how large the gap is in real-world social networks.
Here we address this question by analyzing
100 anonymized networks from the Facebook100 dataset
(see \cite{traud2012social,altenburger2018monophily} for details on this data). This is a useful dataset because with 100 disconnected networks it provides multiple opportunities to test our phenomenon.

Almost all nodes in this data report a gender of male or female,
and so we choose the male and female genders to be our two types
for purposes of measuring homophily. We designate female as one type and male as the other.

To be able to work with the networks, we have to
delete nodes that didn't have a specified gender along with their
connections. Across the 100 networks, 0.92 of nodes on average have a
specified gender (min 0.86, s.d. 0.018). We also want to compute all the 4 different gaps for the same exact sets of nodes, and to be able to look at the singular version of the female \textit{and} the male gap, we need all the nodes in each network to have at least one friend of \textit{each} type.
We prune our networks to only include such nodes. To do this, we remove all the non-compliant nodes, then check if any remaining nodes need to
be removed now that their friends may be gone. A network with no
non-compliant nodes would be done in 1 pass over the nodes (a programming cycle),
and if there are any non-compliant nodes the algorithm would need at
least 2. 6 of the 100 networks were done in 2 cycles, 56 in 3, 34 in
4, and 1 in 5 cycles. We recalculate homophily as the proportion of same-type friends in the new networks. 3 institutions are designated female-only
institutions, and their gendered nodes are $>98.6\%$ one gender, so
pruning leaves very few nodes (0.05, 0.11, and 0.31 of gendered
nodes are left---the next smallest proportion for a co-ed school is 0.89).
We exclude these schools from the analysis. In the end, among 97
co-ed schools, 0.96 of gendered nodes on average are left after
pruning (min 0.91, s.d. 0.015).

\begin{figure}[H]
\centering
\includegraphics[width=\textwidth]{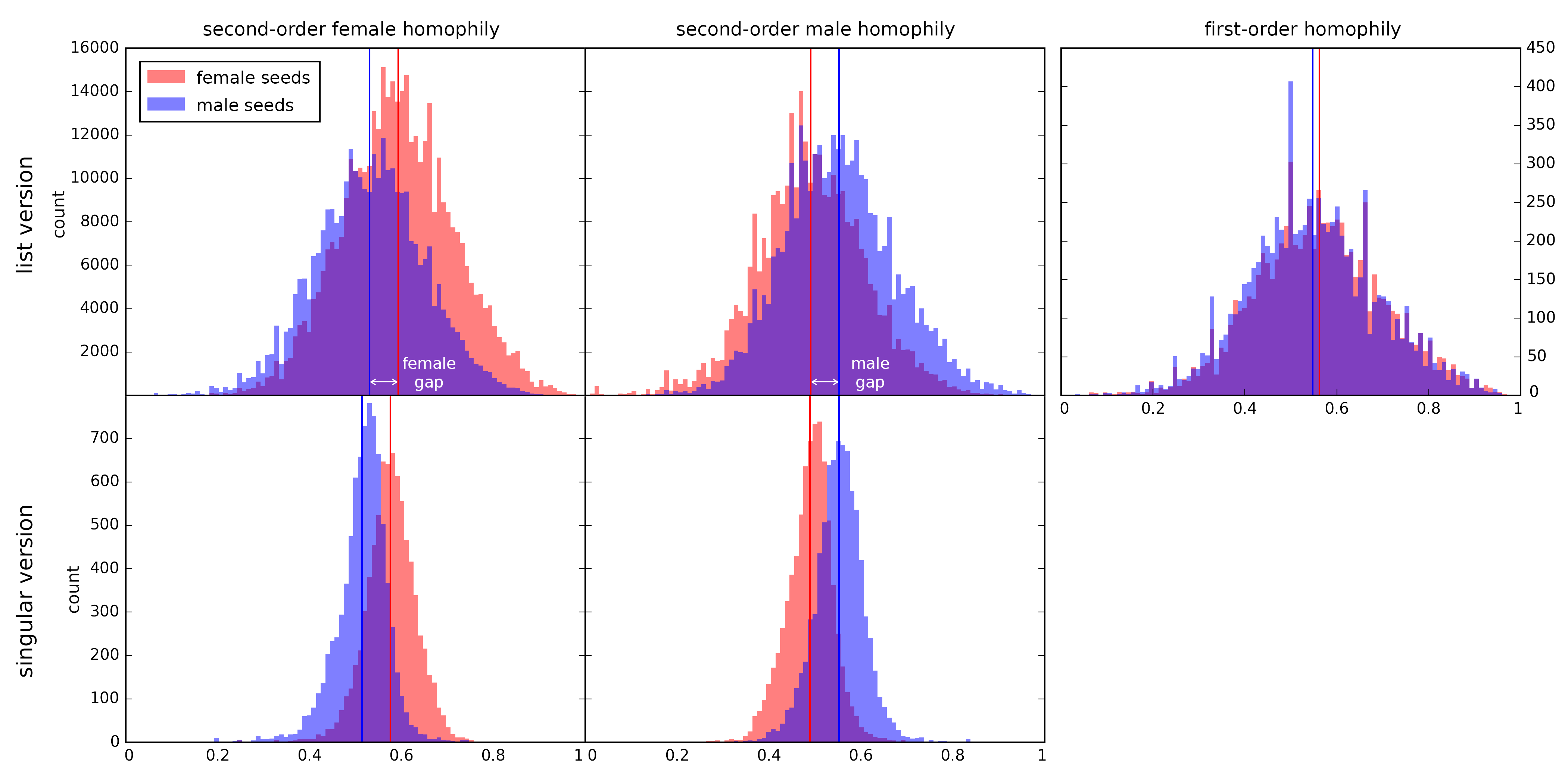}
\caption{Histograms for a co-ed school in the Facebook100 dataset, with gender as type. There are 16822 nodes compliant with our constraints (90.15\% of total nodes), 8687 male and 8135 female. First-order homophily of female and male nodes is reported in the rightmost panel. For the four panels on the left, the top row shows the results for the list version, and the bottom row shows the results for the singular version. The left column shows histograms of second-order female homophily, and the right column shows histograms of second-order male homophily. The gaps for both the second-order female homophily and the second-order male homophily are positive in each version of the phenomenon.
\label{fig:fb100}
}
\end{figure}

We provide a plot for one of the co-ed schools in Figure \ref{fig:fb100}.
When we explore all 97 plots like this, all the female and the male gaps are positive, while the difference between mean first-order homophily of female nodes and mean first-order homophily of male nodes is not always positive or always negative, with a mean of 0.0806 and a standard deviation of 0.0805.

\begin{figure}[H]
\centering
\includegraphics[width=\textwidth]{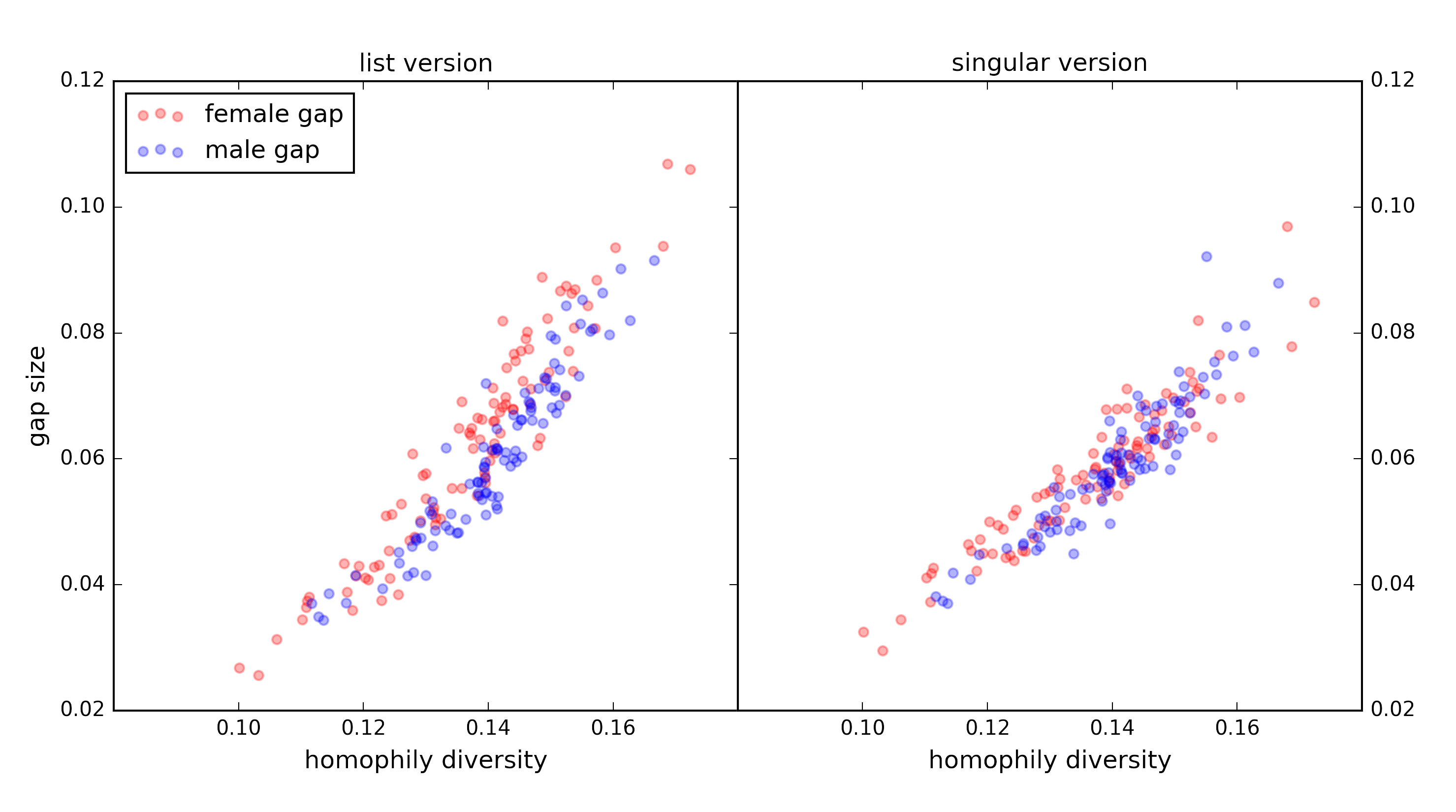}
\caption{For each school whose gap was computed in 
Figure \ref{fig:fb100}, we compute the standard deviation of the first-order homophily values for each type, and plot these against the gap for both the list and singular versions. Consistent with the notion that homophily diversity drives these gap phenomena, we find close alignment between the standard deviation in first-order homophily values and the gap size, with a correlation of $0.94$.
\label{fig:gap-vs-sd}
}
\end{figure}

Given that homophily diversity is a key property in the theoretical formulation of our result, it is natural to ask how large a role the amount of homophily diversity plays in the size of the gap.
We examine this quantitatively in Figure~\ref{fig:gap-vs-sd}, where we plot gap size as a function of homophily diversity (the standard deviation of first-order homophily values) for each type in each of the co-ed schools in our dataset. The first thing to note is the non-negligible gap size. Secondly, we find a remarkably close alignment of gap size with homophily diversity, with a correlation of $0.94$. Additionally, we find it interesting to report that the correlation between gaps for the list and singular versions is $0.903$ for female gaps and $0.92$ for male gaps.

\section{Gap in Random Graphs}

In addition to studying the size of the gap in empirical data,
we can also seek results on the size of the gap in simple random-graph
models of networks with two types of nodes.
In addition to providing a tractable generative model for analysis, 
this approach provides us with a model in which the homophily diversity is directly tunable. In this way, we can see whether the quantitative
dependence of the gap on the amount of homophily diversity, which shows up strongly in the empirical data in Figure~\ref{fig:gap-vs-sd}, is borne out by the model as well.

For this random-graph analysis, we adapt the standard configuration model for random graphs
\cite{newman-sirev} to the case of two types to specifically declare the nodes' degrees and homophily values, or, equivalently, the numbers of friends of each type that each node has. We describe the resulting \textit{double configuration model} in Section 3 of the Supplementary Information. We give an overview of the analysis here,
and provide full details in the Supplementary Information. We also discuss other simulation results there.

{Our main finding here is that it} is possible to arrive at a closed-form formula for the \textit{list version} gap size in networks where first-order homophily of nodes of each type is distributed normally (the sample derived from $\mathcal{N}({\lambda},\,\sigma^{2})$ would be restricted to $[0,1]$ which may lead to deviations in sample mean and s.d.---we refer to this as clipping) and degrees are the same within type ($d$ and $c$ as before). The formula for the list version red gap would be given by 
\begin{equation}
{g^{(R)}} = \ddfrac{\sigma_{R}^2}{{\lambda_R}} +\frac{\sigma_{R}^2}{1-{\lambda_R}}
\end{equation}

where ${\lambda_R}$ and $\sigma_{R}$ are parameters from which the first-order homophily of red nodes sample is constructed. (See Section 5 of the Supplementary Information for derivation details). Note that no other asymmetry parameters (like the number of red nodes $k$ as it relates to the total number of nodes $n$) are part of the formula.

\begin{figure}[H]
\centering
\includegraphics[width=\textwidth]{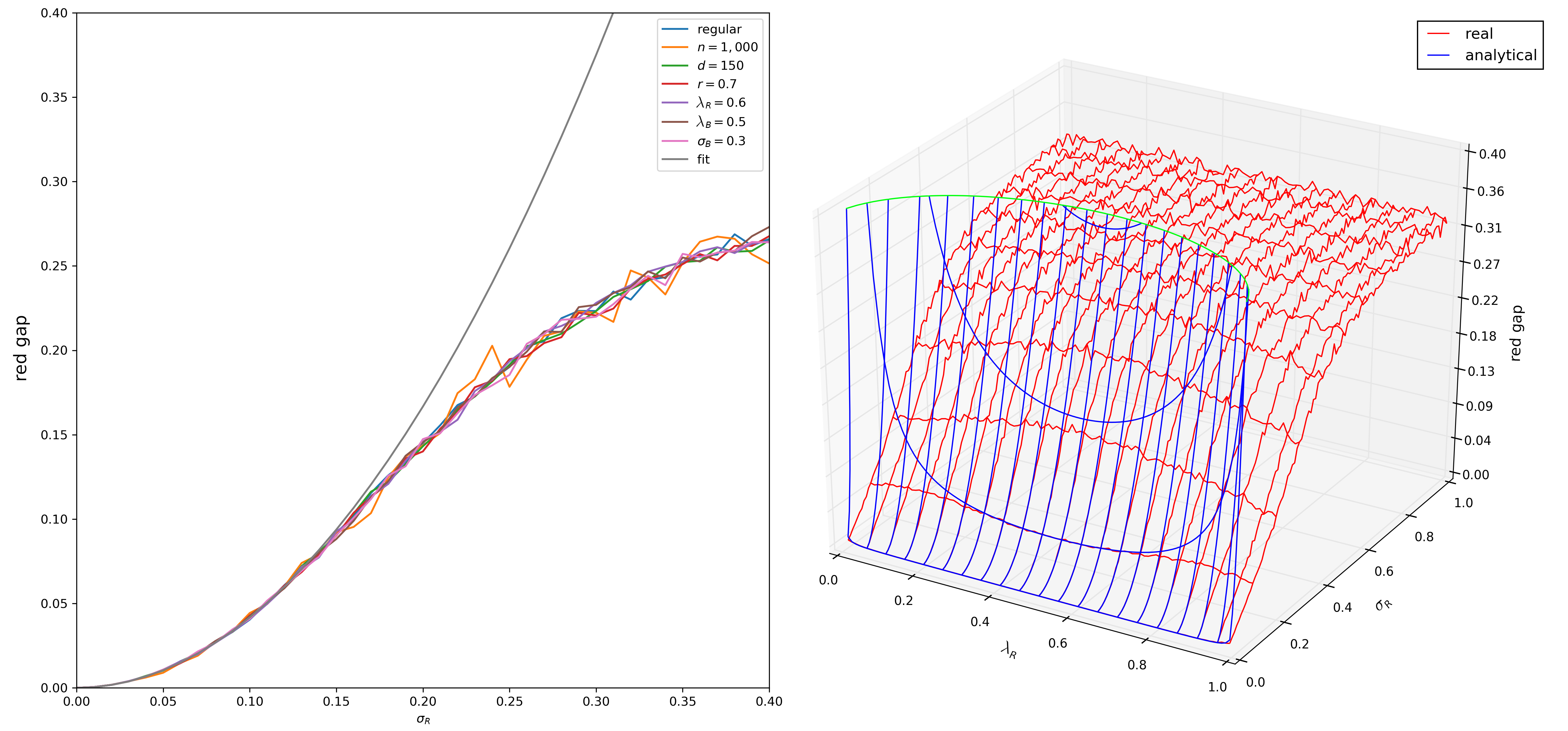}
\caption{\textbf{A:} The parameters used in the regular line are: $n=10000, d=100, r=\frac{k}{n}=0.5, \lambda_R=0.4, \lambda_B=0.3, \sigma_{B}=0.15$. Other simulated lines are the result of changing one of the parameters and intend to show that none of them change the fit of the numerical results to the mathematical prediction. Note that for the ${\lambda_R}$-related line, ${\lambda_R}$ was changed to $1-{\lambda_R}$ which doesn't affect predicted value of the gap given the formula. The prediction fits the simulation well for $\sigma_{R}$ less than roughly $0.125$. The deviations from the prediction stem from clipping effects where the sample is restricted to $[0,1]$ which, as $\sigma_{R}$ increases, make the sample deviate more and more from a true sample from $\mathcal{N}({\lambda_R},\,\sigma_{R}^{2})$, as well as the fact that the gap is necessarily no greater than 1 while the prediction formula has no such constraints. \textbf{B:} $n=10000,d=100,r=0.5,{\lambda_B}=0.3,\sigma_{B}=0.15$. The fit is good for ${\lambda_R}$ which is sufficiently far from 0 and 1. For those extremes, $\sigma_{R}$ is not very representative because the sample ends up being one-sided, clipped from either the $0-$ or the $1+$ side. This shows that red gap does not depend on anything but $\sigma_{R}$ and ${\lambda_R}$, and in simulated results ${\lambda_R}$ has a much lower role than in the prediction. The green line is where the parabola crosses the predicted gap$=0.4$ plane. \label{fig:normal}}
\end{figure}

It is possible to set $n, k, d, \lambda_R, \lambda_B$ and $\sigma_{B}$ and, using the double configuration model, calculate $g^{(R)}$ analytically and numerically to see how it responds to changes in homophily diversity in red nodes $\sigma_R$. Note: once we set all those values, $c$ will be determined; this comes from the balance equation (see equation~\ref{eqn:balance}). Figure~\ref{fig:normal}-A shows simulation results for set parameters with varying $\sigma_{R}$. It also shows how these results respond to changes in the set parameters and how well simulation fits prediction. One of the parameters we vary is the proportion of red nodes $r=\frac{k}{n}$ which we changed from 0.5 to 0.7 (affecting k) to separate the proportion of red nodes as a parameter from two absolute numbers $k$ and $n$. It's clear that changing parameters other than $\lambda_R$ and $\sigma_{R}$ leads to little change in simulated $g^{(R)}$, and results follow the prediction quite closely in the beginning. Later, clipping effects and the fact that $g^{(R)}$ is necessarily limited while the predictive parabola is not lead to discrepancies between the simulation and the prediction.

Since ${\lambda_R}$ and $\sigma_{R}$ are the only parameters influencing $g^{(R)}$, we can see how $g^{(R)}$ responds to the full space of changes in the two parameters of interest with a 3D plot. Note: here again we will set parameters as before and derive $c$ from those---to see how $c$ responds, consult the Supplementary Information. Figure~\ref{fig:normal}-B extends the results of Figure~\ref{fig:normal}-A to similar conclusions. As we reach the extremes of ${\lambda_R}$, clipping effects and no limit on the parabola influence the fit of prediction to simulated results again.

\section{Relationship to the Friendship Paradox}

Given the results presented here, it is useful to return to the 
comparison with the Friendship Paradox that we discussed in the beginning of the paper. The Friendship Paradox, through our lens, looks like this: each node has a first-order number of friends and a second-order number of friends, which is a list or a mean of the first-order numbers of friends of the node's friends. There are no types, and the phenomenon is expressed as the relationship between the first- and the second-order quantities, namely: if there is diversity in the first-order number of friends (not all degrees are equal), the mean second-order number of friends is larger than the mean first-order number of friends for both the list and the singular versions. In order to compute the second-order number of friends for each node we require that each node has at least one friend. Although this definition is similar to our gap phenomenon statement, it concerns the relationship of a second-order quantity to a first order quantity for a network of one type. 

There are instances when the mean second-order red homophily of red nodes is higher than the mean first-order homophily of red nodes, for example, when all red degrees are the same, as shown in the proof of the special case of the singular version (see reasoning below equation~\ref{eqn:singular}). (In fact, this result is a direct consequence of the fact that a subgraph of red nodes would exhibit the Friendship Paradox as there would be degree diversity (with node $i$ having a degree $d h_i$)). However, this is not always the case, as shown in Figure~\ref{fig:example}. While there is both homophily and degree diversity in that network overall and among red nodes specifically, there is no difference in the mean second-order red homophily of red nodes $\mu^{(R)}_{R}$ (or, equivalently in this case, $\mu^{(R\textrm{,sing})}_{R}$) and the mean first-order homophily of red nodes $\lambda_{R}$, giving a gap of 0.

The gap can also be negative in a more complicated case where, for
example, two red nodes with homophily $\frac{1}{2}$ are connected to a
red node with homophily $\frac{1}{100}$ (and degree 200). The list
version of the mean second-order red homophily of red nodes
would be
$(\frac{1}{2}+\frac{1}{2}+\frac{1}{100}+\frac{1}{100})/4=\frac{51}{200}$,
the singular version of the mean second-order red homophily of red nodes would be
$(\frac{1}{2}+\frac{1}{100}+\frac{1}{100})/3=\frac{13}{75}$, and the
mean first-order homophily of red nodes would be
$(\frac{1}{2}+\frac{1}{2}+\frac{1}{100})/3=\frac{101}{300}$ which is
larger than both of the quantities above.

Interestingly, a negative gap in this formulation doesn't only occur
in constructed examples; there is one co-ed school in the Facebook100 dataset where the female second$-$first gap is in fact negative and one other co-ed school where the male second$-$first gap is negative.

Note also that the paradoxes don't have to co-occur. If there is
degree diversity, the Friendship Paradox occurs, but degree diversity
does not guarantee homophily diversity in any type. The opposite---homophily diversity in one or both types with no degree diversity in
the network---is also possible. See Section 6 in the Supplementary Information for examples.

We note that empirically, there is generally a relatively low correlation between the degree of a node and its first-order homophily. Specifically, in the Facebook100 networks the correlation between degree and first-order homophily is generally between -0.25 and 0.2.
This is notable in the context of our gap phenomenon, since it shows another way in which the result for homophily is distinct from the Friendship Paradox for degree: in general, analogs of the Friendship Paradox have held only for close correlates of degree (like level of social activity), but homophily is not a close correlate (nor is our phenomenon a direct extension of the Friendship Paradox).

\section{Possible Generalizations and Applications}

In contrast to our intuition, the gap in second-order homophily means is not due to community structure or the many asymmetries that social and other networks exhibit---all that is required is diversity in first-order homophily of the type where we want to observe a positive gap. In fact, even if the network is \textit{heterophilous}---meaning every node has more friends of the opposite type than of their own---the result holds. In terms of applicability, the phenomenon can be observed in most real-world or simulated networks where two types of nodes can be identified, regardless of their interpretation. The types can be categorically different---such as gender or major party affiliation---or can refer to a binary variable outcome, for example, voting \textit{for} a platform or testing positive for a disease. 

{
The list and singular versions of the phenomenon both imply properties of homophily in network data, but with different emphases in that they are statements about the edges and nodes respectively. 

In particular, the list version of the phenomenon is useful when we have access to random edges of a graph. For example, on social media or messaging platforms, pairs of users will have long-running chat histories, and we can think of these as corresponding to edges between these pairs of users. If the users are from two types (red and blue), then the list version tells us that in comparing random chats between two red nodes (a red-red chat) to random chats between a blue user and a red one (a blue-red chat), we can expect a random red \say{end} of the red-red chat to be more homophilous than the red end of the blue-red chat. This has implications for the inferences we make about homophily when we are sampling pairs based on their communication in this way.

The singular version is useful when we have access to random nodes of a graph, for example, in an idealized case of a field network experiment. Suppose the two types are immune and non-immune and we want to distribute a limited vaccine supply. Selecting a random non-immune friend of a non-immune person would be better than selecting a random non-immune friend of an immune person, as the former will in expectation be more homophilous, meaning proportionally more connected to other non-immune people who will benefit from having an immunized friend. Note that here, because we are interested in the non-immune gap, we would require for all members of the network to have a non-immune friend. However, in practice we don’t have to throw people out of the village, but instead can simply disqualify a random non-immune seed that we select but that happens to have no non-immune friends, and move on to the next one, until we find a suitable seed.}

An interesting direction for further work is to generalize the
definitions to the case of networks with more than two types of nodes. A simple option is to reduce multiple types to two types by focusing on a single type $A$ of interest, and declaring the other type as all nodes not in $A$.

Given the universality of the gap phenomenon we find for two types,
we believe the results here are relevant not just to network theory,
but to any empirical study where it is important to take into account
biases in the data arising from choices of definitions and measurements.

\section*{Supplementary information}
The supplementary information is available via \textit{Scientific Reports} at\\ \underline{\href{https://www.nature.com/articles/s41598-021-92719-6}{nature.com/articles/s41598-021-92719-6}}.

\section*{Data availability} 

The dataset we've used, Facebook100, is introduced in a paper by Traud et al. and is available from the authors of that work~\cite{traud2012social}.

\section*{Acknowledgements} 

We thank Austin Benson, Will Hobbs, and Sigal Oren for their valuable comments on the paper. This work has been supported in part by a Simons Investigator Award and a grant from the MacArthur Foundation. 

\end{document}